\documentclass[]{spie}
\usepackage[]{graphicx}
\usepackage[pdftex]{color}
\usepackage{amsmath}
\usepackage{multirow}

\title{Impact of common modes correlations and time sampling on the total noise of a H2RG near-IR detector}
\author{Bogna Kubik\supit{a}, Remi Barbier\supit{a}, Alain Castera\supit{a}, Eric Chabanat\supit{a}, Sylvain Ferriol\supit{a}, \\
Gerard Smadja\supit{a}
\skiplinehalf
\supit{a} Universit\'e de Lyon, Universit\'e Lyon 1, Lyon F-69003, France \\
CNRS/IN2P3, Institut de Physique Nucl\'eaire de Lyon, Villeurbanne F-69622, France
}

\authorinfo{Further author information: (Send correspondence to B. Kubik)\\
B.Kubik: E-mail: bkubik@ipnl.in2p3.fr, Telephone: +33 4 72 43 10 74\\
G.Smadja: E-mail: g.smadja@ipnl.in2p3.fr, Telephone: +33 4 72 44 83 18 \\
E.Chabanat: E-mail: e.chabanat@ipnl.in2p3.fr, Telephone: +33 4 72 43 27 51\\
S.Ferriol: E-mail: s.ferriol@ipnl.in2p3.fr, Telephone: +33 4 72 44 82 88 \\
A.Castera: E-mail: a.castera@ipnl.in2p3.fr, Telephone: +33 4 72 44 84 29 \\
R.Barbier: E-mail: r.barbier@ipnl.in2p3.fr, Telephone: +33 4 72 43 12 22\\
}

\begin{document} 
\maketitle 


\begin{abstract}
We present the readout noise reduction methods and the 1/f noise response of an 2K$\times$2K HgCdTe detector similar to the detectors that will be used in the Near Infrared Spectrometer Photometer - one of the instruments of the future ESA mission named Euclid.
Various algorithms of common modes subtraction are defined and compared. We show that the readout noise can be lowered by 60\% using properly the references provided within the array. A predictive model of the 1/f noise with a given frequency power spectrum is defined and compared to data taken in a wide range of sampling frequencies. In view of this model the definition of {\it ad-hoc} readout noises for different sampling can be avoided. 
\end{abstract}

\keywords{HgCdTe detectors, H2RG, common modes, reference pixels, 1/f noise, temporal noise, frequency power spectrum, infrared astronomy}

\section{INTRODUCTION}
\label{sec:intro} 

Complementary Metal-Oxide Semiconductor (CMOS) sensors have become a competitive astronomical ground- and space-based detector solution. In the near-infrared spectral range from 1 to 12 $\mu$m, two semiconductor materials are competing, namely InSb and HgCdTe grown by liquid phase epitaxy or molecular beam epitaxy on Al, Si or CdZnTe substrates. This paper will focus on HgCdTe arrays grown on CdZnTe substrates. Hybrid CMOS sensors have a flexible readout structure that allows a group of pixels on the array to be read out at any time without disturbing or reading out the rest of the array. In addition to the flexible readout, CMOS sensors are naturally less sensitive to radiation than more traditional detectors like Charge Coupled Devices (CCDs), since damage to one pixel in the array does not adversely affect subsequent pixels in a row or column of the array. This inherent radiation hardness is particularly appealing for space-based applications. 

An infrared hybrid active pixel sensor is composed of two components, a diode array fabricated with narrow band gap semiconductor material sensitive at infrared wavelengths and a silicon CMOS readout multiplexer with a buffer source follower, a reset switch and addressing switches placed in the unit cell of each pixel. The two components, the infrared diode array and the Si CMOS multiplexer are hybridized with indium bumps making the electrical interconnections for each pixel between the infrared diode and the unit cell of the silicon multiplexer. 

We present here an analysis of the noise of a substrate-removed HAWAII-2RG (HgCdTe Astronomy Wide Area Infrared Imager with 2K x 2K resolution, Reference pixels and Guide mode) array from Teledyne Imaging Sensors (TIS) with a cut-off wavelength of 2.5 $\mu$m and an operational temperature of 90K. The array is driven and digitized by the SIDECAR ASIC chip inside the dewar. The firmware used during acquisitions is the one delivered with the SIDECAR Acquisition Module (SAM Board development kit Software 3.0) provided by the same company \cite{H2RG_tech_man}.

In addition to the common modes investigations through a study of the spatial properties of the noise we implement here a time analysis in each pixel which provides a direct evidence for the 1/f contributions originating from the electronic components at the pixel and readout levels, namely from the PN junction and from the source follower of the multiplexer. The two approaches are complementary, as the spatial analysis is best suited for a characterization of common modes and has led us to propositions for the improvement of the standard acquisition schemes, whilst the time analysis helps to clarify the role of the sampling frequency and has a predictive power of the readout noise in different time sampling.

The work we present in this paper was initiated in the context of the Euclid satellite project. The HgCdTe detectors, usually H1RG 1K$\times$1K, however has been used extensively in numerous focal plane arrays of ground and space telescopes. We shall only quote a small subset of ground implementations in the Wide-field Infrared Survey Explorer (WISE), Very Large Telescope (VLT),  Wide Field Camera 3 (WFC3) of the Hubble Space Telescope (HST). The H2RG-SIDECAR detector system is also planned to be used in the James Webb Space Telescope (JWST).

\section{SETUP}
\label{sec:setup}

The H2RG detector is designed to operate in a number of different modes. We operate with the 100 kHz (slow) pixel rate which is designed to ensure the lowest readout noise. The H2RG detector can be read using 1, 4 or 32 parallel outputs. The former uses the lowest power, but the latter provides the highest frame rates. For the spatial noise analysis we set the 32-output readout mode. With overheads, the acquisition firmware yields a minimum exposure time of 1.47 seconds, or a frame rate of 0.68 Hz, when using 32 channels.

The flexible readout structure allows a sub-array of pixels to be read out at any time, however this mode is only available using one output. In order to achieve very high frame rates for the temporal noise analysis we chose a sub-array of pixels of dimension 8$\times$64 pixels. The 100 kHz pixel rate and some overheads lead to a 7.12 msec minimal exposure time (140 Hz frame rate).

Multiple Accumulated (MACC) sampling and Up-The-Ramp (UTR) sampling are both available. The MACC sampling consist in reading $N$ groups of $n$ frames separated with $n_d$ dropped frames. The $n$ frames in each group are then averaged to form a single coadded frame. In the UTR mode frames are readout continuously during the whole exposure time, without any drops nor coadding. We choose to operate in the UTR mode storing all the frames in the memory and to construct a MACC mode subsequently during the data analysis. We checked that the noise properties are the same in both, the MACC acquisitions programmed directly in the ASIC or the MACC mode constructed in the offline analysis. Our choice of operating only with the UTR acquisitions is driven by the time economy since with one UTR exposure we can test several MACC modes with equivalent exposure time.

Numerous tests were carried out on engineering-grade detector to develop our software (DAS v.1.0 \cite{DAS} and Selene v.1.0 \cite{Selene}) and to characterize the performance of the H2RG-GH456-IR25-GLS SCA. 
The measurements described below were carried out in a dedicated setup built to evaluate the H2RG detectors in the IPNL laboratory facility.

\newpage
\section{COMMON MODE CORRECTION USING REFERENCE PIXELS}
\label{sec:common_modes_ref_pixels}

\subsection{Reference signals}
The sensitivity of infrared detectors is limited by photon shot noise, dark current and read noise. The dark current can be lowered down below the natural background level (such as zodiacal light) in the high quality HgCdTe detectors by cooling them to temperatures below 100 K. The sensitivity of the array is then limited by the read noise. This read noise is mostly the noise of the FETs used to read out the detector, including the statistical noise of the readout FET as well as any noise associated with bias supplies and clocks. While the statistical noise of each unit cell FET is independent, any noise arising from common FETs in the signal chain or from common biases and clocks will be correlated.

The reference signals (that is any signal that electronically mimic the normal pixel output but is not sensitive to light) provided within the detector allow to reduce this common modes at least partially. There are two kind of references provided with the H2RG array, the reference output and the reference pixels.

The sensitive pixel array itself is surrounded by 4 rows and columns of reference pixels included in the array of 2048$\times$2048 pixels. Each of the the detector channels has four row of reference pixels at the top and at the bottom of the array. Additionally the two outside channels have four columns of reference pixels at the outside edge of the array providing a reference for the output at the beginning and at the end of each 64-pixel row.
Reference pixels are not connected to the detector photodiodes but they contain a simple capacitor $C_{pix}$ with capacitance similar to the active pixels capacitance $C_{pix} = 40$ fF. The reference pixels are important to track biases and temperature variations over long exposures and they allow subtraction of the frame to frame bias and temperature drift. 

The reference output is a separate video output read with the same biases that the channels used to sample the active area. The reference signal for the separate reference output is derived from a single pixel with a capacitor of $C_{pix} = 40$ fF. It is read simultaneously and at the same pixel rate that the other outputs. It is subject to the same bias and clock voltage variations as the active channels and can thus be used to correct the signal for these high frequency variations. In the {\it differential} readout of an H2RG array each video output is fed to one side of a corresponding video pre-amplifier in the SIDECAR. The other side of the video output sees the shared reference output signal. The difference of those two signals is sent as the data pixel in 32 data streams. Anther way of reading data is the so-call {\it single ended} readout. In this case the raw signal values on the 32 active-pixel outputs and the reference output are sent in 33 parallel digital data streams with a common 
reference voltage as the second input of the preamplifiers. 

As our aim is to find the optimal correction using reference pixels we do not want to combine the effects of reference output and reference pixels. We operated in the single ended mode where all the pixels are sampled with the same reference voltage $V_{\textrm{refmain}}$ \cite{H2RG_tech_man}. We will compare the best results obtained in this configuration to the noise performances obtained while operating in the differential mode. 

\subsection{Definitions of corrections}
\begin{figure}[ht!]
    \begin{center}
        \begin{tabular}{c}
            \includegraphics[height=7cm]{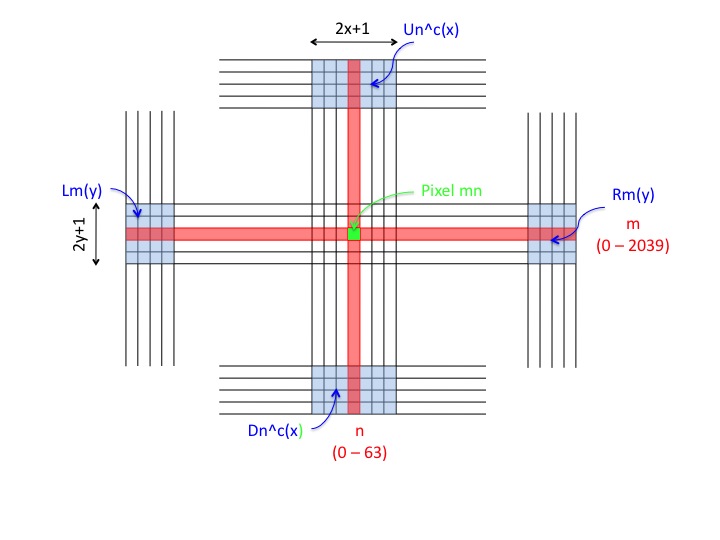}
        \end{tabular}
    \end{center}
    \caption[example] 
   {\label{fig:running_box} The sliding windows parameters definition for a pixel ($m$, $n$).}
\end{figure}
The corrections are performed by subtracting the sliding window averages. We consider a photosensitive pixel $p_{mn}^{(ch)}$ located in the $m-$th line and a $n-$th column of a channel $ch$. For this pixel we define the sliding windows of reference pixels 
\begin{equation}\label{eq:UDLR}
\begin{gathered}
    U_n^{(ch)}(x) = \frac{1}{4(2x+1)}\sum_{i=1}^{4}\sum_{j=n-x}^{n+x} u_{ij}^{(ch)} \, ,\\
    D_n^{(ch)}(x) = \frac{1}{4(2x+1)}\sum_{i=1}^{4}\sum_{j=n-x}^{n+x} d_{ij}^{(ch)} \, ,\\
    L_m(y) = \frac{1}{4(2y+1)}\sum_{i=m-y}^{m+y}\sum_{j=1}^{4} l_{ij} \, ,\\
    R_m(y) = \frac{1}{4(2y+1)}\sum_{i=m-y}^{m+y}\sum_{j=1}^{4} r_{ij}, 
\end{gathered}
\end{equation}
where $x$ and $y$ are the correction parameters which will be adjusted to minimize the noise. The sliding box geometry is presented in Figure (\ref{fig:running_box}). 

The four boxes $U_n^{(ch)}(x)$, $D_n^{(ch)}(x)$, $L_m(y)$ and $R_m(y)$ defined in Equation (\ref{eq:UDLR}) can be combined to form the corrections $c_1$ to $c_3$ as follows\footnote{Note that this is not an exhaustive list of linear combinations of $U_n^{(ch)}(x)$, $D_n^{(ch)}(x)$, $L_m(y)$ and $R_m(y)$. We have checked that the cases listed in this paper are the most representative ones to illustrate effects interesting for us. This list includes the most effective correction to mitigate common modes effects.}:
\begin{equation}\label{eq:corrections} 
\begin{gathered}
\begin{aligned}
    c_{1n}^{(ch)}(x)    &= \frac{1}{2}\left[U_n^{(ch)}(x) + D_n^{(ch)}(x)\right]\, , \\
    c_{1mn}^{(ch)}(x)   &= D_n^{(ch)}(x) + (n-1.5)\tan\alpha_n^{(ch)}\, , \\
    c_{2m}(y)           &= \frac{1}{2}\left[L_m(y) + R_m(y)\right] \, ,\\
    c_{2mn}(y)          &= L_m(y) + (n-1.5)\tan\beta_m \, ,\\
    c_{3mn}^{(ch)}(x,y) &= \frac{1}{2}\left[U_n^{(ch)}(x) + D_n^{(ch)}(x)\right]  + \frac{1}{2}\left[ L_m^{'out}(y) + R_m^{'out}(y) \right] \, ,
    \end{aligned}
    \end{gathered}
\end{equation}
with
\begin{equation}
\begin{gathered}
\begin{aligned}
    \tan\alpha_n^{(ch)} & = \frac{U_n^{(ch)}(x) - D_n^{(ch)}(x)}{2044}\, ,\\
    \tan\beta_m         & = \frac{R_m(y) - L_m(y)}{60}\, , \\
    L'_m(y)             & = \frac{1}{4(2y+1)}\sum_{i=m-y}^{m+y}\sum_{j=1}^{4} \left[l_{ij} - c_{1n}^{(0)}(x)  \right] \, ,\\
    R'_m(y)             & = \frac{1}{4(2y+1)}\sum_{i=m-y}^{m+y}\sum_{j=1}^{4} \left[r_{ij} - c_{1n}^{(31)}(x) \right] \, .\\
    \end{aligned}
    \end{gathered}
\end{equation}
The correction $c_1$ uses only the windows of up and down reference pixels, the $c_2$ - the windows of the left and right reference pixels. Both cases come with two variants - either a simple average calculation [$c_{1n}^{(ch)}(x)$ and $c_{2m}(y)$] or with a slope fitting [$c_{1mn}^{(ch)}(x)$ and $c_{2mn}(y)$]. The correction $c_{3mn}(x,y)$ uses both, the up/down and left/right references.
Correcting a photosensitive pixels $p_{mn^{(ch)}}$ means subtracting a value calculated by one of the formulas in Equation (\ref{eq:corrections})
\begin{equation}
    p_{mn}^{(ch)} ~~\longrightarrow~~ p_{mn}^{(ch)} - c_{imn}^{(ch)}(x,y) = p_{mn}^{(ch)\textrm{corrected}} \, .
\end{equation}

\subsection{Data and results}
For the spatial analysis of the noise we take a sample of 100 UTR dark exposures of 100 frames separated by two reset frames. With this sample we construct the CDS and Fowler(16) readout modes using Selene v.1.0 \cite{Selene}. Fowler(16) signal is defined as the mean value of differences between the consecutive groups. The groups are constructed by coadding and averaging 16 frames. We do not insert drops between groups so as to minimize the effects of stray photons and the $1/f$ effects. The standard deviation form the mean value of Fowler(16) signal is the Fowler(16) noise. 

The H2RG detector is cooled down to the operating temperature of 93 K ($\pm10$ mK) with liquid nitrogen and is operated in the unbuffered mode. The measured signal in the absence of any illumination  is found to be of 1 e$^{-}$/sec. In the analysis we treat the total noise including the dark and the stray photon noise. The CDS noise of the raw data is 31 e$^{-}$ r.m.s. and the Fowler(16) noise is measured at the level of 17 e$^{-}$ r.m.s.

\begin{figure}[ht!]
    \begin{center}
        \begin{tabular}{cc}
            \includegraphics[scale=0.3]{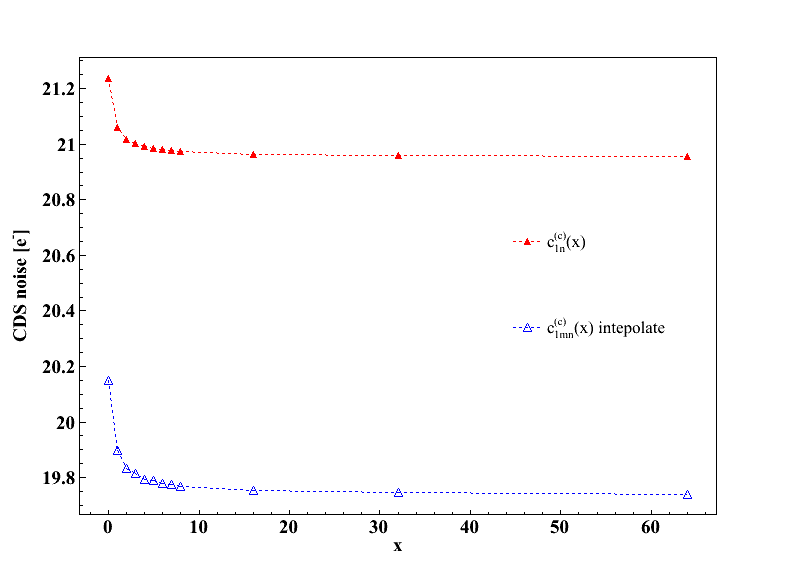}
            \includegraphics[scale=0.3]{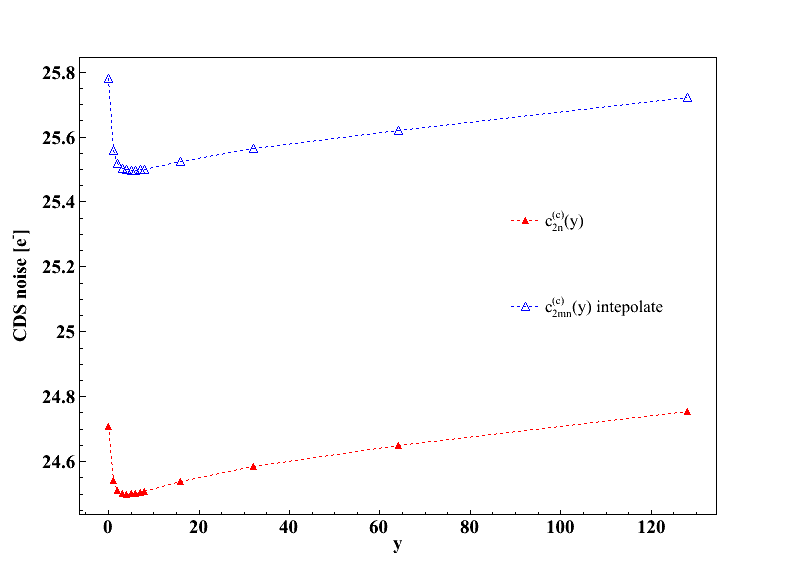}\\
            \includegraphics[scale=0.3]{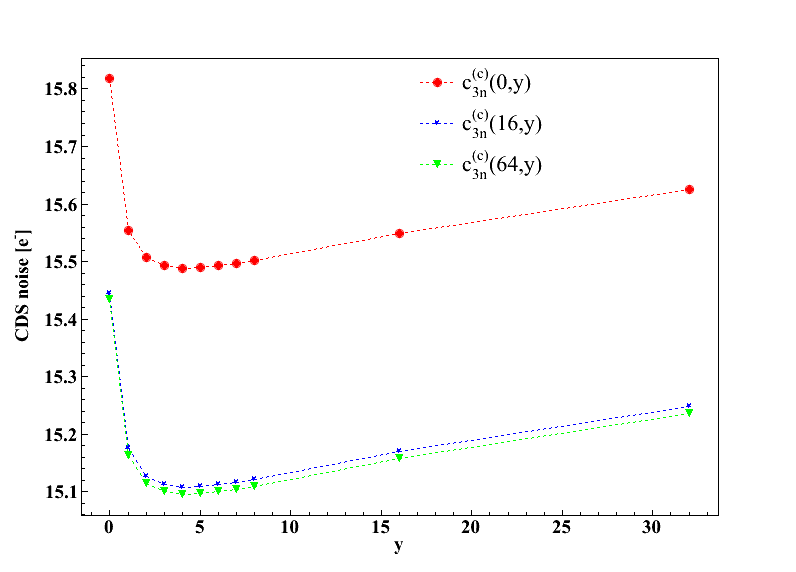}
        \end{tabular}
    \end{center}
    \caption[example] 
   {\label{fig:cds_noise_box_size_dependence} CDS noise in e$^{-}$ r.m.s. in the channel 20 after reference pixels corrections with different window size. Left: Corrections $c_{1n}^{(20)}(x)$ and $c_{1mn}^{(20)}(x)$ interpolated. Middle: Corrections $c_{2m}(y)$ and $c_{2mn}(y)$ interpolated. Right: Corrections $c_{3mn}(x,y)$ for different parameter sets. The most effective correction is $c_{3mn}(x,y)$ with $x=64$, $y=4$.}
\end{figure}

Our analysis, summarized in Figures (\ref{fig:cds_noise_box_size_dependence}) shows that the optimal box for the up and down pixels is the full channel average: $x=64$ with the CDS noise reduced by 0.4 e$^{-}$ r.m.s and the Fowler(16) noise reduced by about 0.03 e$^{-}$ r.m.s. For the left and right reference pixels averages the optimal parameter is $y=4$ pixels. The results imply as well that the up/down averages should be interpolated before being subtracted - improving the CDS by about 1.5 e$^{-}$ r.m.s and the Fowler(16) noise by about 0.1 e$^{-}$ r.m.s. The left/right averages should be subtracted without interpolation. We have checked that in the case of correction $c_{3mn}^{(ch)}(x,y)$ the interpolation effects are less important than in the correction schemes $c_{1mn}^{(ch)}(x),\,c_{2mn}^{(ch)}(y)$ changing the final value of the CDS noise and Fowler(16) noise by less than 0.1\%. 
\begin{figure}[ht!]
    \begin{center}
        \begin{tabular}{cc}
            \includegraphics[scale=0.3]{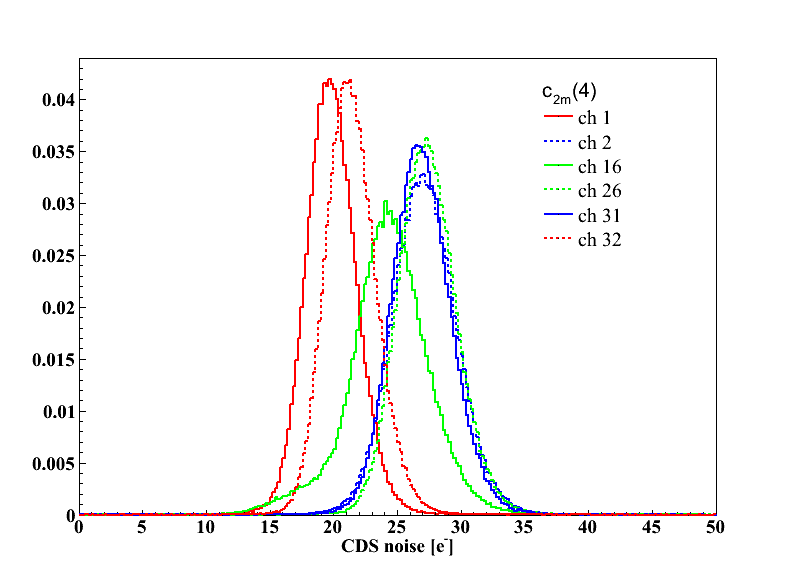}
            \includegraphics[scale=0.3]{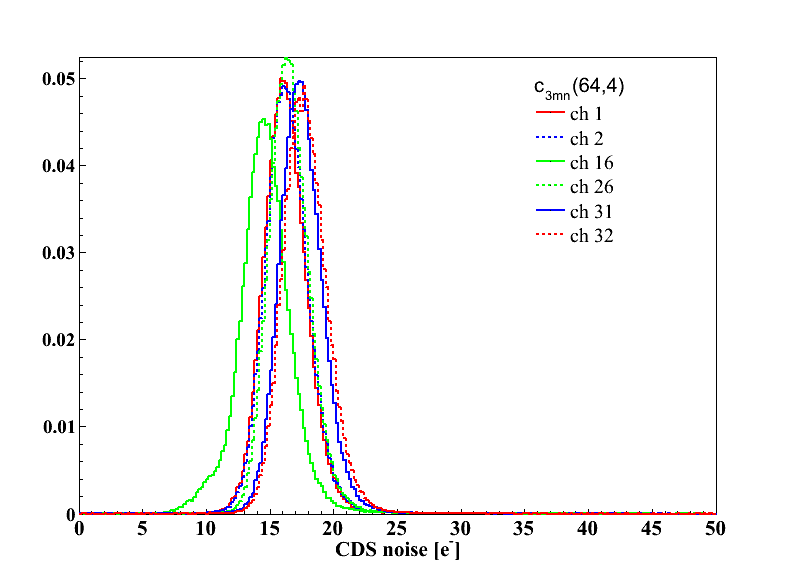}
        \end{tabular}
    \end{center}
    \caption[example] 
   {\label{fig:output_dependence} CDS noise in e$^{-}$ r.m.s. after reference pixels corrections. Left: $c_{2m}(4)$. Right: $c_{3mn}^{(ch)}(64,4)$ for channels 0, 1, 15, 25, 30 and 31. Correction $c_{2m}(4)$, which uses only left and right reference pixels, reduces the noise in the channels 0 and 31 more effectively that in other channels. Correction $c_{3mn}^{(ch)}(64,4)$ uses left/right and up/down reference pixels and reduces the noise to the same level in all the channels independently on their position.}
\end{figure}
As the left/right reference pixels lie only in the two external video channels, the position of the channel being corrected can play a role. It is effectively the case if one uses only left and right reference pixels [corrections $c_{2n}$ and $c_{2mn}$]. The corrections $c_{1m}$, $c_{1mn}$ and $c_{3mn}$ use only up and down reference pixels available in each channel thus there is no reason to introduce a gradient between the noise of different channels as illustrated in Figure (\ref{fig:output_dependence}). A way to decrease the noise homogeneously in all channels using the left/right and up/down is to apply the correction $c_{3mn}^{(ch)}(x,y)$ where the left and right reference pixels are corrected by the up and down averages before being averaged and subtracted from the sensitive pixels.

Our analysis shows (see Table (\ref{tab:cmc_summary}))  that the most effective correction in removing common modes is the correction $c_{3mn}^{(ch)}(64,4)$ which reduces the CDS noise by 51\% and the Fowler(16) noise by 65\%. The spatial RMS of the noise is lowered by 33\% and 21\% for CDS and Fowler(16) noise respectively. For comparison we have also tested the analog and digital subtraction of the reference output. The performances of the common mode corrections with reference pixels in these configurations turned out to be worse, with the minimal CDS noise achieved of about 22 e$^{-}$ r.m.s.
\begin{table}[ht!]
{\footnotesize
    \begin{center}
        \begin{tabular}{ccccc}
            \hline
            cor&                         CDS &        spatial & Fowler(16) &        spatial \\
               &                        noise&        RMS     & noise      &        RMS     \\
            \hline
            
              raw            &           30.9&           2.70 & 17.4       &       1.53 \\
             $c_{1n}(64)$    &           21.0&           2.15 &  7.1       &       1.120 \\ 
             $c_{2m}(4)$     &           24.5&           2.31 & 8.2        &       1.17 \\ 
             $c_{1mn}(64)$   &           19.7&           2.33 & 7.1        &       1.19 \\
             $c_{2mn}(4)$    &           25.5&           2.56 & 8.5        &       1.23 \\
             $c_{3mn}(64,4)$ &           15.1&           1.82 & 6.1        &       1.22 \\
            \hline
        \end{tabular}
        \caption{\label{tab:cmc_summary}CDS and Fowler(16) noise in $e^-$ r.m.s. for the 500 mV polarized H2RG detector and different common modes corrections. The spatial RMS over a 2040 $\times$ 64 output indicates the homogeneity of the noise repartition. Both, the best noise reduction and homogeneity over the array are obtained for the $c_{3mn}(64,4)$ correction.}
    \end{center}
    }
\end{table}
\newpage
\section{RESET NOISE AND TEMPERATURE DRIFTS MITIGATION WITH REFERENCE PIXELS}

For the reset fluctuation study we have taken 400 exposures of two frames. The detector substrate voltage $D_{\textrm{sub}}$ is set to a constant value of 550 mV and we change the detector reset voltage $V_{\textrm{reset}}$ from 50 mV to 350 mV with steps of 50 mV at four operating temperatures set to 90 K, 95 K, 100 K and 105 K. We compute the reset noise per pixel as the standard deviation form the averaged reset level over all the 400 exposures taken at a given temperature and given $V_{\textrm{reset}}$. The values reported below are the spatial averages over the array of $64 \times 2040$ pixels that belong to one video output. The variations of the reset level as a function of the temperature or $V_{\textrm{reset}}$ are also computed at pixel level and the spatial averages are reported in this paper. 

\begin{figure}[ht!]
    \begin{center}
        \begin{tabular}{cc}
            \includegraphics[scale=0.3]{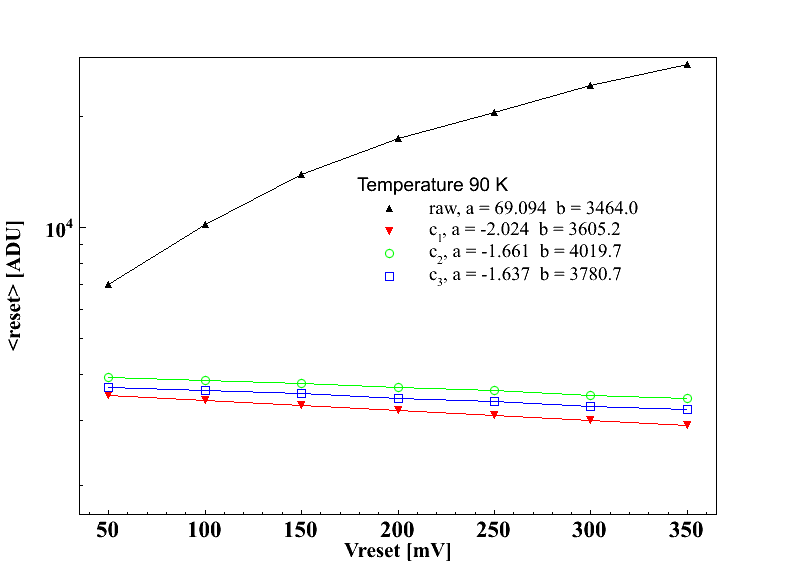}
            \includegraphics[scale=0.3]{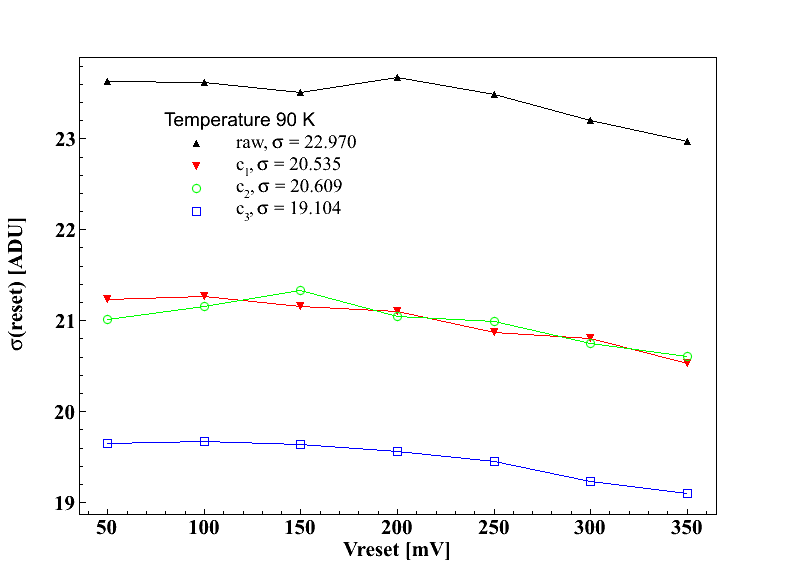}\\
            \includegraphics[scale=0.3]{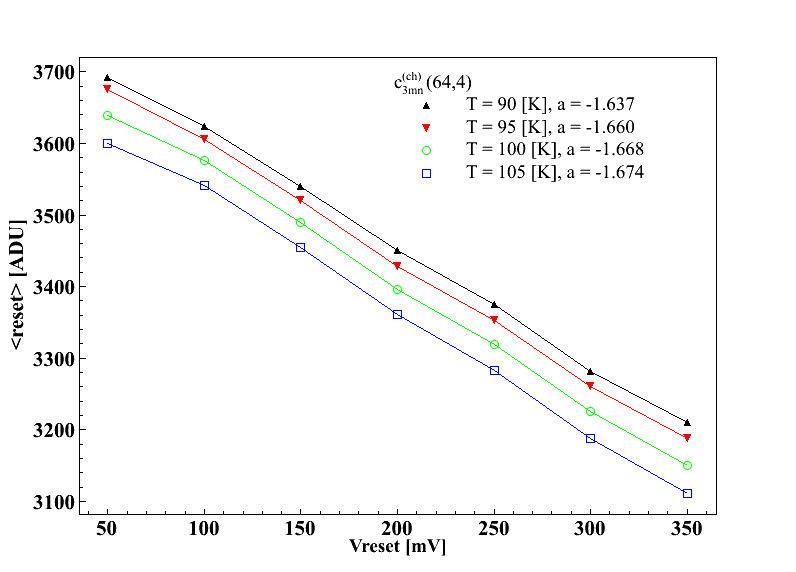}
            \includegraphics[scale=0.3]{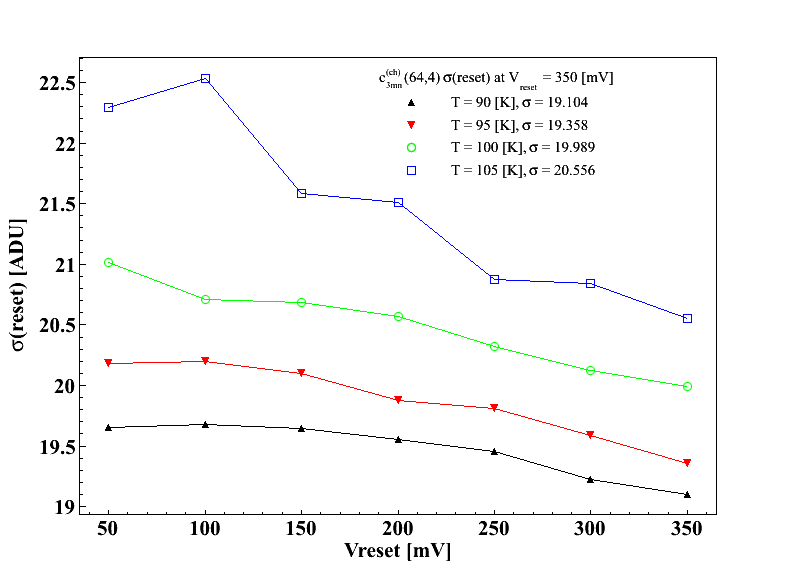}
        \end{tabular}
    \end{center}
    \caption[example] 
   {\label{fig:reset_vs_Vreset} On the top: reset level and reset noise per pixel as function of $V_{\textrm{reset}}$ for different reference pixels corrections at the operating temperature 90 K. Correction $c_{3mn}$ is the most effective in removing the $V_{\textrm{reset}}$ drifts and minimizes the reset noise  On the bottom: reset level and reset noise variations at different operating temperatures but with the same common modes removal scheme. The slope $a$ [ADU/mV] which gives the correction that should be applied to completely mitigate the $V_{\textrm{reset}}$ drifts increases slightly with the temperature in the temperature rage 90 - 105 K. The reset noise increases with the operating temperature as expected, but decreases with $V_{\textrm{reset}}$ value.}
\end{figure}

As can be seen in Figure (\ref{fig:reset_vs_Vreset}) and in Table (\ref{tab:reset_vs_Vreset}) the $V_{\textrm{reset}}$ drifts can be easily reduced with the reference pixels which are subject to almost the same variations with reset level as the photosensitive ones. The 70 ADU/mV drift is effectively reduced to only 1.6 ADU/mV. The correction for $V_{\textrm{reset}}$ modulation (the value of the slope [ADU/mV]) increases with the operating temperature as illustrated in the bottom left panel of Figure (\ref{fig:reset_vs_Vreset}). The minimal reset noise, which increase with the operating temperature as expected, is obtained in the $350$ mV polarization of the substrate and using the $c_{3mn}$ correction. With the reference pixels correction the reset noise is lowered by more than 15\% compared to its initial value for raw data. 

\begin{figure}[ht!]
    \begin{center}
        \begin{tabular}{cc}
            \includegraphics[scale=0.3]{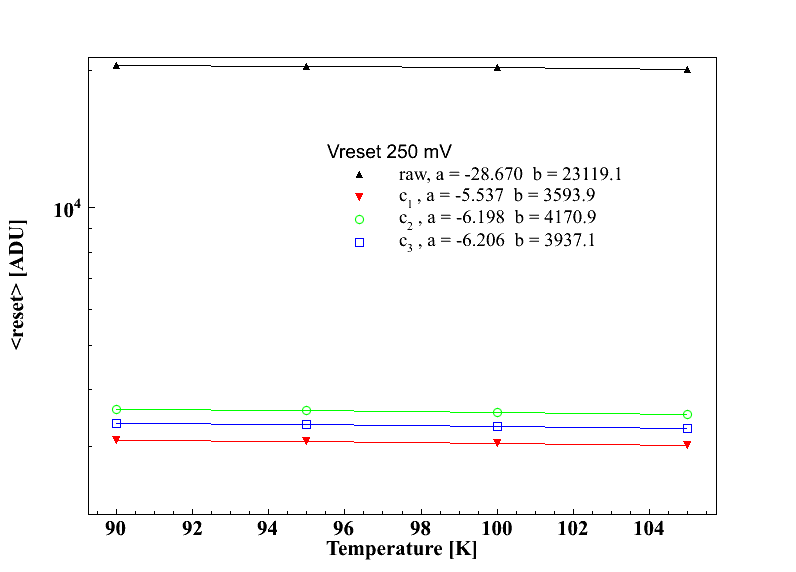}
            \includegraphics[scale=0.3]{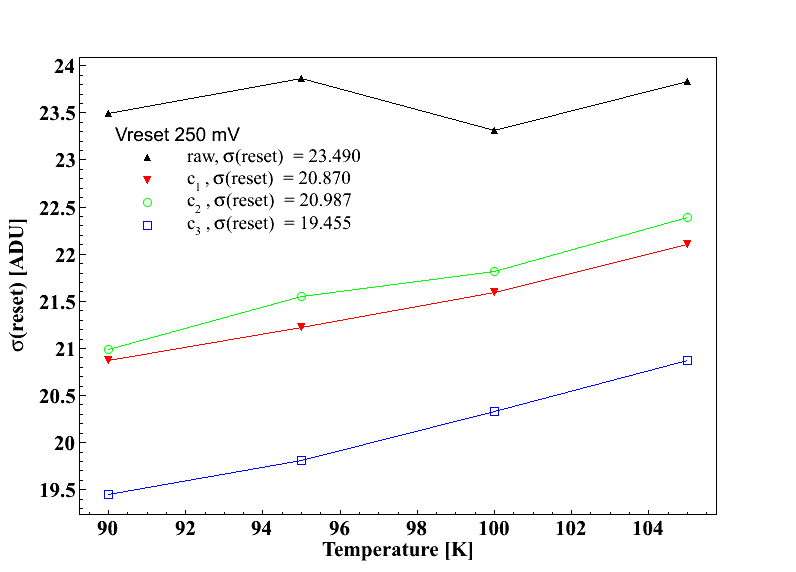}\\
            \includegraphics[scale=0.3]{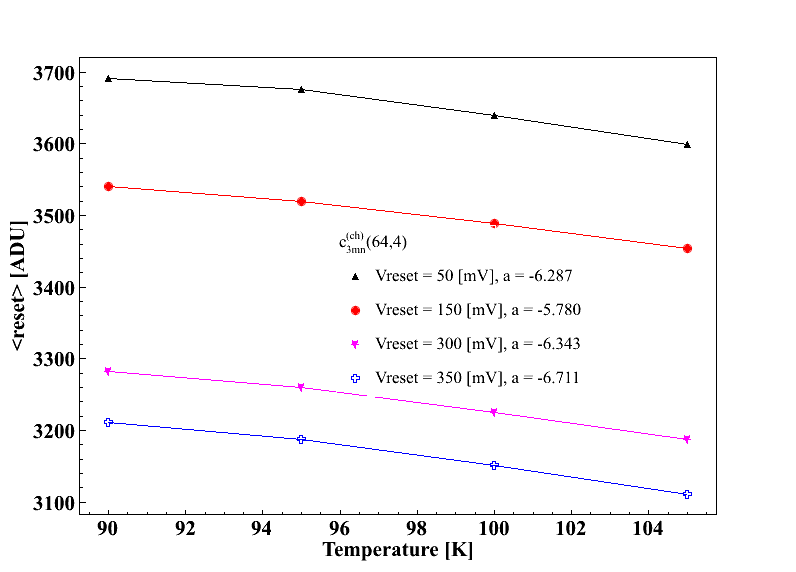}
            \includegraphics[scale=0.3]{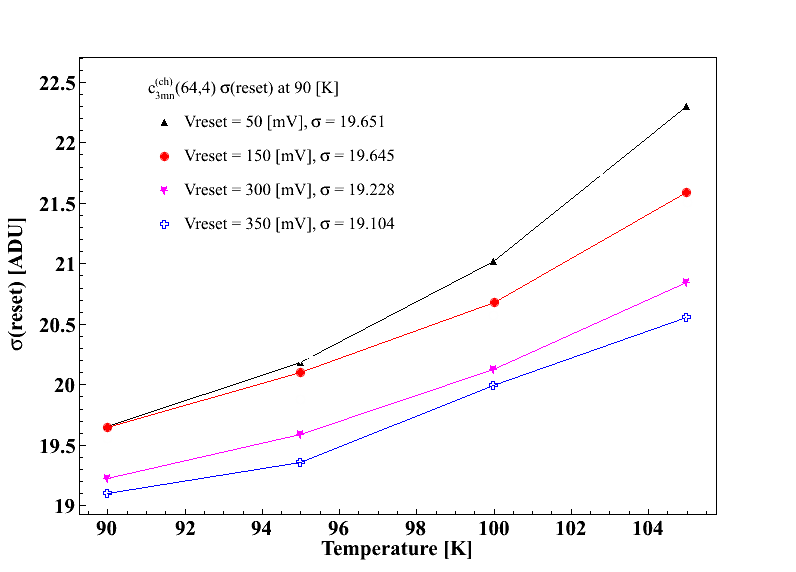}
        \end{tabular}
    \end{center}
    \caption[example] 
   {\label{fig:reset_vs_T} On the top: reset level and reset noise per pixel as function of the temperature for different reference pixels corrections for $V_{\textrm{reset}}$ = 250 mV. We note an increase in the reset noise with the operating temperature and the correction $c_{3mn}$ removes partially the noise of the first frames. On the bottom: Variation of the reset level and reset noise as a function of the operating temperature and for different $V_{\textrm{reset}}$ values after applying the $c_{3mn}$ common mode correction. There is no visible trend in the slope $a$ [ADU/K] while the reset noise decrease with the substrate polarization.}
\end{figure}

The temperature drifts mitigation is also a very important issue as any not controlled temperature variation may induce a large noise increase through dark noise and reset noise modification. The reference pixels can easily remove this unwanted effect at least partially as illustrated in Figure (\ref{fig:reset_vs_T}) and in Table (\ref{tab:reset_vs_T}) where the raw pixel variation with temperature is reduced by a factor of 5. Also the reset noise is reduced by 20\% at the operating temperature of 90 K.

\begin{table}[ht!]
{\footnotesize
    \begin{center}
        \begin{tabular}{c|cc|cc|cc|cc}
            \hline
            {\bf Temp}     &{\bf 90 K}&             &{\bf 95 K}&             &{\bf 100 K}&            &{\bf 105 K}&  \\
            \hline
            {\it corr}     & $a$    & r.m.s         & $a$    & r.m.s         & $a$    & r.m.s         & $a$    & r.m.s \\
                           & ADU/mV &(350 mV)       & ADU/mV &(350 mV)       & ADU/mV &(350 mV)       & ADU/mV &(350 mV)\\
            \hline
            {\it raw}      & 69.094 & 22.970        & 68.991 &  23.293       & 69.008 & 23.206        & 69.043 & 23.436         \\
            $c_{1mn}(64)$  & -2.024 & 20.535        & -2.047 &  20.90        & -2.060 & 21.335        & -2.078 & 21.703         \\
            $c_{2mn}(4)$   & -1.661 & 20.609        & -1.679 &  20.879       & -1.690 & 21.654        & -1.692 & 22.085         \\
            $c_{3mn}(64,4)$& -1.637 & 19.104        & -1.660 &  19.358       & -1.668 & 19.989        & -1.674 & 20.556         \\
            \hline
        \end{tabular}
        \caption{\label{tab:reset_vs_Vreset}Reset level variation with $V_{\textrm{reset}}$ (the slope $a$ [ADU/mV] is reported) for different reference pixels corrections and at four different operating temperatures. The slope $a$ increases slightly with the temperature. The reset noise per pixel (in [ADU r.m.s.]) increases with the operating temperature as expected, but decreases with $V_{\textrm{reset}}$ value. In the table for each temperature only the minimal value of reset noise (reached at $V_{\textrm{reset}}$ = 350 mV) is given.}
    \end{center}
    }
\end{table}

\begin{table}[ht!]
{\footnotesize
    \begin{center}
        \begin{tabular}{c|cc|cc|cc}
            \hline
            {\bf $V_{\textrm{reset}}$}& {\bf 50 mV}     &           & {\bf 150 mV}        &           & {\bf 250 mV}    & \\ 
            \hline
            {\it corr}     & $a$       & r.m.s.        & $a$       & r.m.s.        & $a$       & r.m.s.   \\
                           & [ADU/K]   & (90 K)        & [ADU/K]   & (90 K)        & [ADU/K]   & (90 K)  \\
            \hline    
            {\it raw}      & -28.759   & 23.628        & -28.852   & 23.505        & -28.670   & 23.49 \\
            $c_{1mn}(64)$  & -4.809    & 21.239        & -5.155    & 21.152        & -5.537    & 20.87 \\
            $c_{2mn}(4)$   & -6.516    & 21.019        & -5.658    & 21.330        & -6.198    & 20.99 \\
            $c_{3mn}(64,4)$& -6.287    & 19.651        & -5.780    & 19.645        & -6.206    & 19.46 \\
            \hline
        \end{tabular}
        \caption{\label{tab:reset_vs_T}Reset level variation with operating temperature  (the slope $a$ [ADU/K] is reported) for different reference pixels corrections and at different $V_{\textrm{reset}}$ levels. The slope $a$ decreases slightly with $V_{\textrm{reset}}$. For each $V_{\textrm{reset}}$ the minimal value of reset noise per pixel (reached at T = 90 K) is given.}
    \end{center}
    }
\end{table}

\newpage 
\section{TEMPORAL NOISE ANALYSIS}
\label{sec:common_modes_window}
For the temporal analysis we select in the 2K $\times$ 2K H2RG detector array a window of 8$\times$64 pixels. The time needed to read the entire 8$\times$64 pixels window is $\delta$ = 7.12 msec (140 Hz frame rate) with the 100 kHz pixel rate and the overheads. We take 25 non-destructive exposures UTR of 200000 frames that is a total exposure time of 1424 sec. The selection of a small window allows to reach a higher repetition rate for the frame readout and to save on the overall measurement time as well as the on computing resources. In the analysis we eliminate first 50 frames in each exposure to avoid any initial instability after the reset. The remaining frames are combined into $N$ groups $G_k$ of $n$ frames $s( t )$ forming MACC modes with different group periodicities $\Delta$. The time $t$ is the time when a pixel is readout and $t_0$ is the time of the last reset of the given pixel before the beginning of the exposure. In this analysis, as we are interested only in differences between consecutive 
groups, the $t_0$ 
constant 
can be set to 0 without loosing any accuracy in the following calculations.

\begin{figure}[ht!]
    \begin{center}
        \begin{tabular}{cc}
            \includegraphics[scale=0.4]{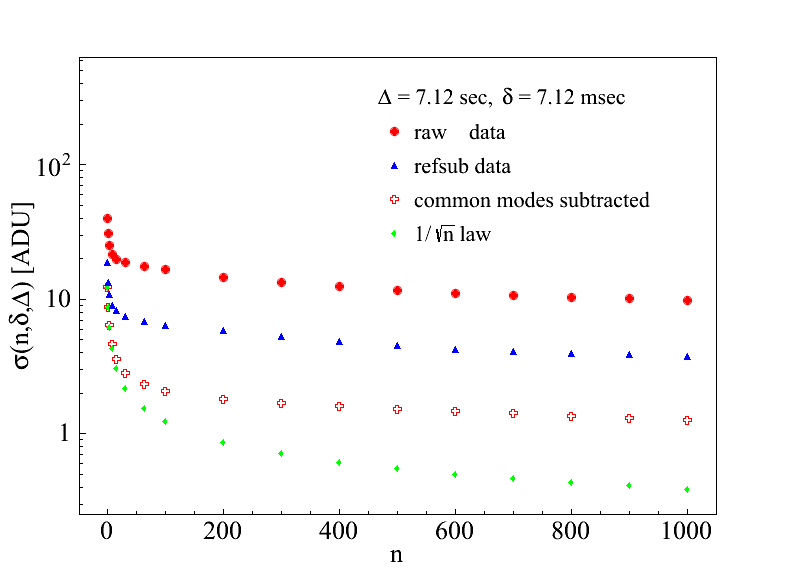}
        \end{tabular}
    \end{center}
    \caption[example] 
   {\label{fig:Fowler_function} Fowler noise $\sigma_F^2(n,\delta,\Delta)$ as a function of $n$. The contribution of the dark current and stray photons has not been subtracted but their contribution is negligible. The noise is computed for raw data, then for pixels after reference channel subtraction (refsub data) and finally after subtraction of common modes by spatial averaging.}
\end{figure}

The Fowler($n$) noise $\sigma_F(n,\delta,\Delta)$ is evaluated in each pixel by the average of $D_k(n,\delta,\Delta)$ over $N$ groups stored during the exposures using relations
\begin{equation}
    D_k(n,\delta,\Delta) = \frac{1}{n}\sum_{i=1}^{n} \left[ s( t_0 + k\Delta + i\delta ) -  s( t_0 + (k-1)\Delta + i\delta ) \right] \, ,
\end{equation}
\begin{equation}
    \sigma_F^2(n,\delta,\Delta) = \frac{1}{N-1} \left[ \sum_{k=1}^{N} \left( D_k - \langle D_k \rangle \right)^2 \right]\, ,
\end{equation}
with 
\begin{equation}
    \langle D_k \rangle = \frac{1}{N} \left( \sum_{k=1}^{N} D_k \right)\, .
\end{equation}
As in the window mode we do not have access to the reference pixels we reduce a part of the common modes by taking the spatial average of each frame and subtracting it from each pixel in the corresponding frame. The contribution of high frequency modes is reduced by subtracting the reference channel. The impact of these common modes corrections is shown in Figure (\ref{fig:Fowler_function}) for an exposure of 195 groups composed of $n$ ranging from 1 to 1000 frames. The group periodicity was set to 1000 frames, i.e. $\Delta = $7.12 sec and the number of dropped frames is adjusted for each $n$ value. We observe that the subtraction of common modes reduces the Fowler noise by more than 50\% for large Fowler samplings ($n \geq 8$).
The value reached for Fowler(1000) corresponds to a Fowler noise of 1.6 ADU. It is apparent that the noise does not decrease like $1/\sqrt{n}$, as would be expected for independent measurements and which would lead to a group to group noise of 0.36 ADU for $n = 1000$ frames. The main contribution to these time correlations arises from the trapping and detrapping of electrons in the readout electronics or in the detector itself. This time correlation has already been noted in several previous publications such as \cite{Finger2004}, \cite{Smith2006} or \cite{Schubnell2006} and can be described by a frequency power spectrum. The connection between the time constant of trapping sites and the frequency power spectrum was first stated in \cite{Schottky1926}.

\begin{figure}[ht!]
    \begin{center}
        \begin{tabular}{cc}
            \includegraphics[scale=0.4]{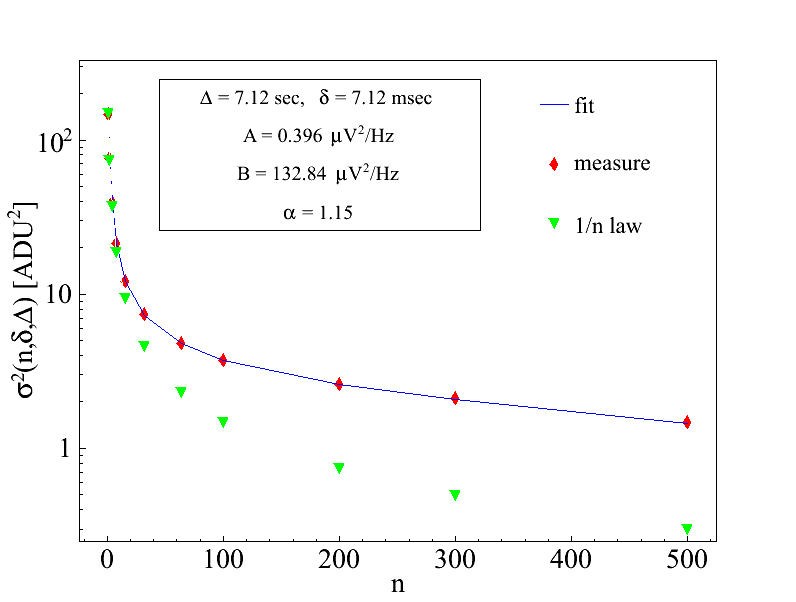}
        \end{tabular}
    \end{center}
    \caption[example] 
   {\label{fig:Fowler_fits} The measured group to group readout noise as a function of the number of
repeated reads and the fitted parameter of the noise power spectrum.}
\end{figure}

The extraction of the frequency power spectrum of the noise is performed as described in \cite{Smadja2010}. The method relies on the Wiener-Khinchin theorem which relates the autocorrelation of a signal to its power spectrum $f (\omega )$ by the relation  
\begin{equation}
    \langle s(t) s(t+\tau) \rangle = \int_{\omega_1}^{\omega_2} \cos ( \omega \tau ) | f (\omega ) |^2 d\omega \, .
\end{equation}
$\omega_1$ and $\omega_2$ are needed to ensure the convergence of the integrals, but are chosen so as to cover all the frequencies occurring in the actual exposure. The explicit expression of the variance written as a sum of autocorrelation products is derived in \cite{Smadja2010}. The measured variance in the absence of stray photons can be expressed as
\begin{equation}\label{eq:Dk}
\begin{gathered}
\begin{aligned}
    \langle D_k^2 \rangle = & \int (1 - \cos (\omega \Delta) ) \left( A + \frac{B}{\omega^{\alpha}} \right) \times \\
    &\left[ \frac{2}{n} + \frac{4}{n}\frac{ \cos(n \omega \delta/2) \sin( (n-1)\omega\delta/2)}{\sin (\omega\delta/2) } 
     + \frac{1-\cos(n\omega\delta) -2\sin((2n-1)\omega\delta/2)\sin(\omega\delta/2)}{n^2\sin^2(\omega\delta/2)} \right].
\end{aligned}
\end{gathered}
\end{equation}
Writing the noise function in this form allows to extract the noise power spectrum function $f(\omega)$ with an accuracy better than 1\% in a wide frequency range and predict the noise behavior for different timing conditions. Following the authors of \cite{Smadja2010} the power spectrum is assumed to be of the form
\begin{equation}
    | f(\omega) |^2 = A + \frac{B}{\omega^\alpha}\, .
\end{equation}
As the authors of \cite{Smadja2012} pointed out, this form of the parametrization is suggested by the general frequency behavior of circuit components. 
To the Equation (\ref{eq:Dk}) should be added the shot noise contribution of any incident photon source subject to
stochastic fluctuations. The impact of such a constant photon source on the total Fowler($n$) variance was derived in \cite{Smadja2010} and, although in a context of a straight line fit, in\cite{Rauscher2007}. The signal contribution $\langle P^2_n \rangle$ to the variance can be written as
\begin{equation}
    \langle P^2_n \rangle = DI + \beta di\, ,
\end{equation}
where $DI$ is the signal between two groups (between the last frame of a group $k$ and the first frame of a group $k+1$) and $di$ is the frame to frame integrated flux in electrons. The factor $\beta$ takes into account the correlations arising from the non-destructive readouts.
\begin{equation}
    \beta = \frac{(n-1)(2n-1)}{3n}
\end{equation}
Figure (\ref{fig:Fowler_fits}) shows the result of an adjustment of the noise in a single exposure with the parameterized power spectrum $\langle D^2_k \rangle + \langle P^2_n \rangle$. To convert the ADU units into $\mu$V /Hz we have used the pixel capacitance of 40 fF. With the frame periodicity $\delta = 7.12 $ msec and the group periodicity $\Delta = 3.56$ sec we find the parameters $A = 0.396$ [$\mu$V$^2$/Hz], $B=132.84$ [$\mu$V$^2$/Hz] and $\alpha = 1.15$. We have also checked the consistence of the fit with a different group periodicities $\Delta = 0.46$ sec and $\Delta = 14.24$ sec maintaining the same frame periodicity $\delta = 7.12$ $\mu$sec. The coefficients $A$ and $B$ turned out to be of the same order, the power exponent $\alpha$ were found to be 1.14 for $\Delta = 0.46$ and 1.34 for $\Delta = 14.24$. The results are summarized in Table (\ref{tab:fit_values}).
\begin{table}[ht!]
{\footnotesize
    \begin{center}
        \begin{tabular}{ccc|ccc}
            \hline
                       & $\delta$ & $\Delta$ & $A$             & $B$             & $\alpha$ \\
                       & [msec]   &  [sec]   & [$\mu$V$^2$/Hz] & [$\mu$V$^2$/Hz] &          \\
            \hline
            {\bf set 1}& 7.12     & 0.46     & 0.388           & 119.04          & 1.14 \\
            {\bf set 2}& 7.12     & 3.56     & 0.396           & 132.84          & 1.15 \\
            {\bf set 3}& 7.12     & 14.24    & 0.412           & 139.12          & 1.34 \\
            \hline
        \end{tabular}
        \caption{\label{tab:fit_values}Adjusted power spectrum parameters for group periodicities $\Delta = 0.456$ sec, 3.56 sec and 14.24 sec.}
    \end{center}
    }
\end{table}

\subsection{Predictions for the Fowler noise for different time sampling}
\label{sec:predictions}
We can also check how accurately the evolution of Fowler noise with the group and frame period is predicted using the frequency power spectrum with the parameters found in our analysis. The contribution of dark and stray photons to the measured and predicted noise is negligible. The white noise component is always found to be dominant. As expected, the variance increases when the time delay between successive Fowler groups increases. Moreover, we observe that the noise decreases with the frame delay $\delta$ increasing. This effect may be due to the correlation range of frames. For small $\delta$ the sampled frames in one group are strongly correlated resulting in the higher noise level than for larger $\delta$. In the latter case, the weaker correlations between frames tend to smooth the signal level in all the Fowler groups, which effectively lowers the Fowler noise. Both effect are confirmed by the predicted and measured values. At the frame rates of 7.12 msec the white noise contribution is significant 
but the contribution of $1/f$ 
effects has a non negligible effect on the outcome of different samplings. We show in Table (\ref{tab:predictions1}) the Fowler(16) variance for different group samplins. In the range $\Delta < 15$ sec the parameters in set 1 and set 2 describe the noise with an accuracy better than 3\%. For lower group frequencies ($ 15 < \Delta < 50$ sec) we should rather use the parameters in set 3, which were  obtained as a result of the fit for $\Delta = 14.24$ sec, in order to reach the accuracy of about 10\%. In the very low frequency regime, $\Delta > 50$ sec, in order to predict accurately the variance the power $\alpha$ must be increased up to 1.5.  We report also the variance values for the time samplings proposed in Euclid mission ($\delta = 1.3$ sec $\Delta = 24$ or $37$ sec). The predicted values for the total variance reproduce the measured ones with the precision better than 10\%.
 \begin{table}[ht!]
{\footnotesize
    \begin{center}
        \begin{tabular}{ccccc}
        \hline
        $\delta$ [sec] & $\Delta$ [sec] & \multicolumn{2}{c}{$\sigma_F^2(n,\delta,\Delta)$ [ADU${}^2$]} & \\ 
                       &                & predicted              &measured& \\                   
        \hline
        0.00712 & 0.57           & 10.28                         & 10.23 & \multirow{3}{*}{{\bf set 1,2}} \\
                & 7.40           & 12.36                         & 12.66 &\\
                & 14.70          & 13.38                         & 13.81 &\\
        \hline
        0.00712 & 21.50          & 16.42                         & 15.17 & \multirow{2}{*}{{\bf set 3}} \\
                & 35.71          & 18.52                         & 21.96 &\\
        \hline
        0.00712 & 57.07          & 26.10                         & 26.17 & \multirow{3}{*}{{\bf $\alpha = 1.5$}} \\
                & 71.31          & 28.47                         & 29.16 &\\
                & 106.91         & 33.82                         & 37.98 &\\
        \hline
        0.1     & 20             & 16.56                         & 16.56 & \multirow{2}{*}{{\bf $\alpha = 1.5$}} \\ 
        1       &                & 15.18                         & 14.14 & \\
        \hline
        1.5     & 24             & 12.39                         & 13.52 & \multirow{2}{*}{{\bf $\alpha = 1.5$}} \\ 
        1.5     & 37             & 15.15                         & 14.53 & \\
        \hline
        \end{tabular}
        \caption{\label{tab:predictions1}Predictions of Fowler noise for different frame samplings $\delta$ and group samplings $\Delta$. The contribution of dark and stray photons to the measured noise is negligible. The white noise component is always found to be dominant.}
    \end{center}
    }
\end{table}

\newpage
\section{CONCLUSIONS}
We have defined and tested various common mode corrections using reference pixels. We have shown that both, the vertical and horizontal reference pixels should be used simultaneously in order to reduce the CDS noise by 51\% and the Fowler(16) noise by 65\%.  We have also demonstrated that the same common mode correction has the advantage of reducing strongly the reset noise and the reset level variations due to drifts of biases and temperature, which might be present during the mission. 
It was found that the reset noise per pixel is reduced by 20\% at the operating temperature of 90 K. The drifts of photosensitive pixels  while the reset level varies in time are effectively reduced by 95\% (the 70 ADU/mV drift is effectively reduced to only 1.6 ADU/mV) and the variation of photosensitive pixels with temperature is reduced by a factor of 5 with the common mode correction $c_{3mn}$.

The temporal noise analysis has led us to the prediction of the frequency power spectrum. The power spectrum of the noise can be described by three parameters to account quantitatively  for the Fowler($n$) variation of the noise at different group periodicities within an accuracy of the order of 3\%.  The power law exponent $\alpha$ found to reproduce the data varies in the range between 1.14 and 1.5.  
The parametrization of the power spectrum depends only mildly on the actual cycling times implemented in the measurements, in contrast to the empirical parametrizations. The evolution of the noise can be predicted for different sampling schemes and the observed variances are in good agreement with the predicted values. The decrease of the variance in the same data set but with increasing frame sampling is an evidence for $1/f$ contributions to the noise in the low frequency range.

\acknowledgments
The authors would like to thank the engineers form the Institut de Physique Nucl\'eaire de Lyon that have contributed to the development of the cryogenic facility for their efforts. The authors thank also the support from the CNRS/IN2P3.
This research was conducted within the framework of the Lyon Institute of Origins under grant ANR-10-LABX-66.
\bibliographystyle{spiebib}

\end{document}